\documentclass[11pt]{elsart}
\usepackage[dvips]{graphicx}

\begin{document}
\begin{frontmatter}

\title{Interplay between HIV/AIDS Epidemics and Demographic Structures
Based on Sexual Contact Networks
%\footnote{This work has been
%supported by the Chinese Natural Science Foundation of China under
%Grant Nos. 70431001 and 70271046.}
}

\author{Wen-Jie Bai$^{a,b}$ }
\author{Tao Zhou$^{a}$ }
\author{and Bing-Hong Wang$^{a}$}

\address{$^{a}$ Department of
Modern Physics, University of Science and Technology of China, Hefei
Anhui, 230026, P R China}
\address{$^{b}$ Department of
Chemistry, University of Science and Technology of China, Hefei
Anhui, 230026, P R China}

\begin{abstract} \textnormal{\small {In this article, we propose a network spread model for HIV
epidemics, wherein each individual is represented by a node of the
transmission network and the edges are the connections between
individuals along which the infection may spread. The sexual
activity of each individual, measured by its degree, is not
homogeneous but obeys a power-law distribution. Due to the
heterogeneity of activity, the infection can persistently exist at a
very low prevalence, which has been observed in real data but can
not be illuminated by previous models with homogeneous mixing
hypothesis. Furthermore, the model displays a clear picture of
hierarchical spread: In the early stage the infection is adhered to
these high-risk persons, and then, diffuses toward low-risk
population. The prediction results show that the development of
epidemics can be roughly categorized into three patterns for
different countries, and the pattern of a given country is mainly
determined by the average sex-activity and transmission probability
per sexual partner. In most cases, the effect of HIV epidemics on
demographic structure is very small. However, for some extremely
countries, like Botswana, the number of sex-active people can be
depressed to nearly a half by AIDS. }}

\end{abstract}
\date{}
\end{frontmatter}

\section{Introduction}
AIDS (Acquired Immure Deficiency Syndrome), as one of the most
dangerous diseases over human history, has been continuously
spreading at an enormous speed with an extremely high rate of
death (from the moment when the first infection case was
confirmed). Now it has already spread to all the regions of the
world and been a great threat not only to the human health but
also to the human society due to its own epidemiologic characters
which make the objection to AIDS an extremely complex and
difficult task to address.

A high infection rate of population will cause catastrophe to the
development of national economy for two reasons. On one hand, most
infected people are in a group aged from 24 to 45 who do the main
contributions to the country's productivity, thus the AIDS will
cause a great decline of the social wealth. On the other hand, to
carry out a widely covered treatment program on treating the HIV
infected people will be a heavy burden for the government finance
due to the great cost on expensive medicine as well as regularly and
continuously implemented therapy. Therefore, to investigate the
edpidemic behaviors of HIV/AIDS appears not only of great
theoretical interest in understanding the underlying spreading
mechanism, but also necessary and urgent in practice.

The extensively investigated models for epidemics, such as the
standard susceptible-infected-removed (SIR) and
susceptible-infected-susceptible (SIS) models, often involve two
hypotheses \cite{Class1,Class2,Class3,Class4}. First, the population
is assumed closed, that is, the population size is fixed. However,
recent researches on the spread of HIV (especially in Africa and
other worst-afflicted areas) indicate the existence of strongly
interplay between HIV epidemics and age structures, thus the
demographic impact can not be neglected (see the review papers
\cite{Review1,Review2} and the references therein). Secondly, the
epidemiological models are often established based on perfect and
homogeneous mixing, that is to say, all individuals are able to
infect all others and the infectivity of each individual is almost
the same. To replace the perfect mixing assumption, one can
introduce the \emph{epidemic contact network}, wherein the nodes
denote individuals and edges represent the connections between
individuals along which the infection may spread (see the review
papers \cite{Review3,Review4} about network epidemics and the
references therein). The infected individual can infect a
susceptible one only if they are neighboring in the network. The
homogeneous mixing assumption can be implemented by using epidemic
contact networks with homogeneous degree distributions
\cite{Grassberger1983}, such as regular lattices, random networks
\cite{Bollobas1985}, and so on. However, recent empirical data
exhibit us that the real-world sexual contact patterns are far
different from the homogeneous ones
\cite{Lilieros2001,Lilieros2003,Schneeberger2004}. The corresponding
networks, similar to many other real-life networks, display the
so-called scale-free property \cite{Barabasi1999,Albert2002}, that
is, they are of power-law degree distributions. This power-law
distribution falls off much more gradually than an exponential one,
allowing a few nodes of very large degree to exist. These
high-degree nodes are called \emph{hub nodes} in network science
\cite{Albert2002,Review5,Review6,Review7} and \emph{superspreaders}
in epidemiological literatures \cite{Bassetti2005,Small2005}. Recent
theoretical researches on epidemics show that the topology of
epidemic contact networks will highly affect the dynamical behaviors
\cite{Pastor-Satorras2001,May2001}, and it is also demonstrated that
the effect of the superspreaders on HIV epidemics can not be ignored
\cite{Hyman2001,Hyman2003}. Therefore, the structural effect should
be taken into account when modeling HIV/AIDS epidemics. In addition,
the introduction of Antiretroviral (ARV) drug therapy is also one of
the important result-dependent factors \cite{Blower2000,Blower2003}.
No treatment is more efficient for HIV-infected individuals than the
medicine, which combines two or three antiretroviral drugs in
"cocktail" regimens. These regimens, known as highly active ARV
therapy, have resulted in the reduction of HIV levels in the blood,
often to undetectable levels, and have markedly improved immune
function of HIV-infected individuals. The advent and widespread
application of ARV has dramatically changed the typical course of
HIV infection and AIDS, especially in high-income countries
\cite{Porco2004}. On the other hand, however, in the low-income
countries, the overwhelming proportion of HIV-infected persons has
no access to ARV. In sub-Saharan Africa, for example, this lack of
treatment access has transformed into rapidly escalating death
rates. Although the usage of ARV appears effective in bating HIV, it
will bring too heavy pressure in economy for poor countries.
Therefore, the better understanding of the effect of ARV treatment
may enlighten readers in allocating the financial resources.

Although the demographic structure
\cite{Anderson1988,Surasiengsunk1998} and sexual contact pattern
\cite{Rothenberg1998,Potterat1999}, respectively, has been taken
into account in the previous HIV/AIDS epidemic models, there are few
works simultaneously consider these two ingredients. In the present
model, both the demographic impact and heterogeneity mixing effect
are considered. And many important features of real-life HIV
epidemics can be naturally generated by combining these two
ingredients. This article is organized as follows: In section 2, the
model is presented in details. In section 3, the main properties of
this model are shown. Then, in section 4, we will try to predict the
HIV/AIDS epidemics by this model. Finally, we sum up this article
and discuss the relevance of this model to the real world in section
5.

\section{Model}

\subsection{Construction of Epidemic Contact Networks}
The HIV is transmitted by body fluid through several main routes
including sexual contacts, sharing injectors among drug users,
perinatal transmissions, transfusion of contaminated blood products
etc., which are closely related to human beings' social activities.
In different regions the popularity of each route is variable
according to the culture and social circumstance. In some areas the
homosexual contacts and injecting drug use play the main role in HIV
epidemics, while the main track in HIV transmission is the
heterosexual contacts in the global scope \cite{UNAIDS}. Therefore,
in this model, only the heterosexual relationships are taken into
account, thus the corresponding epidemic contact networks are
bipartite graphs \cite{Ergun2002,Holme2003}.

The epidemic contact network starts with $N_0/2$ males and $N_0/2$
females, each of which is sex-active with age between 15 and 49,
and only the heterosexual contacts are permissive. Since men tend
to over-report their number of partners whereas women tend to
under-report, the total numbers of sexual partners of males and
females are not equal in existing surveys
\cite{Lilieros2001,Surasiengsunk1998}. However, for
simplification, we assume the degree distribution for both male
and female nodes are the same. Assign each male node's degree
according to a given degree distribution $p(k)\sim k^{-\gamma}$
with minimal degree $k_{min}=1$. According to the empirical data
in Sweden \cite{Lilieros2001}, we set $\gamma=3.5$. After
obtaining the \emph{degree sequence} of male nodes, we let the
female nodes have the same degree sequence and randomly assign
each female node's degree according to this sequence. Here the
degree sequence means a set of all nodes' degrees, one can find a
more detailed and strict definition in ref. \cite{Chung2002}. The
edges are generated randomly by using the mechanism of
configuration model \cite{Newman2001}. Note that, different from
most previous studies about epidemics on static networks, the
present network structure evolves with time according to some
followed rules of HIV epidemic dynamics.

\subsection{Modelling spread of HIV/AIDS in networks}
We focus on the network susceptible-infected-removed (SIR) model in
which individuals can be in three discrete states, susceptible,
infected or removed (dead). The infected ones can be divided into
two subclasses: The HIV-positive individuals and persons with AIDS.
Since the median time from AIDS to death is very short (about 7
month for adults \cite{Kitayaporn1996}) compared with the median
incubation time for AIDS \cite{Munoz1997}, we assume that when an
HIV-positive person becomes an AIDS-patient, she or he will
immediately be in death (i. e. within one year). This model is
implemented by computer simulation with a time step equal to one
year when mimicking the reality. The simulation processes are as
follows.

(1) Set all the nodes to be susceptible except one randomly
selected infected one.

(2) At each time step for each susceptible node $x$, denote $m_1$
and $m_2$ the number of its neighboring infected nodes not in
process of ARV treatment (non-ARV user) or contrary (ARV user),
respectively. The probability that the node $x$ will become
infected in the next time step is
\begin{equation}
\pi_x=1-(1-\beta_1)^{m_1}(1-\beta_2)^{m_2},
\end{equation}
if $x$ is male. Here $\beta_\bullet$ is the transmission
probability per sexual partner, which is considered as a more
appropriate estimate than probability per sexual act
\cite{Garnett1994}, and the subscript represents whether the
corresponding HIV-positive person has taken the ARV treatment.
Since the male-to-female transmission is about twice efficient as
female-to-male transmission \cite{Mastro1996}, if $x$ is female,
the corresponding probability is
\begin{equation}
\pi_x=1-(1-2\beta_1)^{m_1}(1-2\beta_2)^{m_2},
\end{equation}
where $\beta_1$ and $\beta_2$ are restricted below 0.5. It has been
estimated by an analysis of longitudinal cohort data that
antiretroviral therapy reduces per-partnership infectivity by as
much as 60\% \cite{Porco2004}, thus we set $\beta_2=0.4\beta_1$.

(3) At each time step, each infected node (except the newly
infected ones) may die with probability either $\zeta_1$ (for
non-ARV user) or $\zeta_2$ (for ARV user). According to the recent
estimations \cite{Sabin2005}, we set $\zeta_1=0.15$ and
$\zeta_2=0.08$. The dead individuals are removed from the
population.

Repeat these processes for desired time. Note that, each newly
infected node will be ARV user at probability $\rho_{\texttt{ARV}}$,
and all the existing ARV users will keep using ARV.

\subsection{Demographic Impact}
In this model, all the nodes (sex-active persons) are divided into
7 age-groups (labeled A1-A7): 15-19, 20-24, 25-29, 30-34, 35-39,
40-44, 45-49. At beginning, each node chooses to be in one
age-group with probability according to the age structure in the
year corresponding to time step zero. At each time step, each
female individual may bear a child according the corresponding
age-specific fertility rates. If she is infected and has not taken
ARV treatment, the perinatal transmission probability is
$\varepsilon_1$. And it reduces to $\varepsilon_2$ if ARV
treatment is taken. Based on some previous empirical studies
\cite{Anderson1988,Chotpitayasunondh1997}, we set
$\varepsilon_1=0.4$ and $\varepsilon_2=0.2$. The infected elder
persons ($>49$ year) may die with probability $\gamma_1$ or
$\gamma_2$ during each time step, and the corresponding
probability for perinatally infected children is 0.2
\cite{Chotpitayasunondh1997}.

At the end of each time step, 1/5 randomly selected living persons
in age-group A1-A6 will reach the elder group, and 1/5 randomly
selected living persons in group A7 will be removed from this
system. If the time step $t$ is less than 15, we simply assume
equal number (to the number of removal nodes in A1) of susceptible
individuals will be added to group A1; else if $t\leq 15$,
$b(t-15)$ individuals will be added to group A1, where $b(t)$
denotes the number of newborn babies without HIV at time $t$. Here
we simply assume all the infected babies will die before 15 years
old since the mortality per year for them is much higher than
adults. All these newly added ones will joint the epidemic contact
network according to the rules of \textbf{Section 2.1}, that is,
the female/male nodes will randomly choose sexual partners among
all the young and old men/women according to their given degrees
that obey the distribution $p(k)$.

See the \textbf{Appendix A} for the source of all the population
and demographic data.

\section{Main Properties}
There are three free parameters in the present model: the average
degree $\langle k \rangle$ which determine the degree distribution
when the power-law exponent $\gamma=3.5$ is given, the
transmission probability $\beta_1$, and the ARV-receiving rate
$\rho_{\texttt{ARV}}$. The former two parameters are relative to
the behaviors while the last one is partially dependent on
financial conditions. In this section, we will show some
simulation results and investigate the main properties about this
model by adjusting the above parameters. Some previous works show
that for most cases, the qualitative features of epidemic dynamics
will not be affected by the slightly varying of population size
and age-structure
\cite{Review1,Anderson1988,Surasiengsunk1998,Nelson1998}, thus in
this section, the age-specific fertility rates are kept unchanged.
We use the age-density and age-specific fertility rates of China
in 2005 for initialization, with the age-specific fertility rates
unchanged all through. The network size is $N=10^6$.

\begin{figure}
  \begin{center}
       \center \includegraphics[width=13cm]{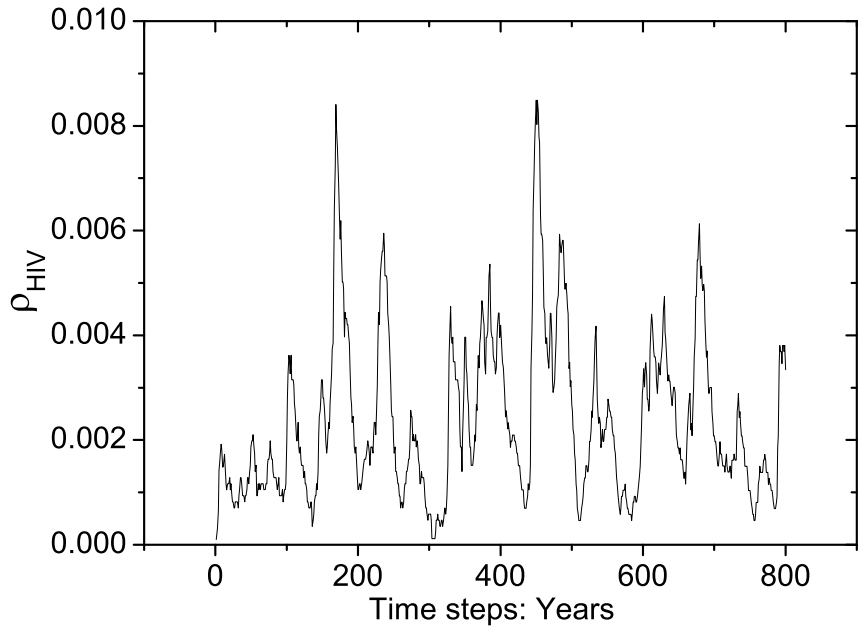}
       \caption{The prevalence of HIV verse time. $\rho_{\texttt{ARV}}$ denotes the ratio of HIV-positive individuals to the whole population of sex-active ones (i. e. the network size). The corresponding
       parameters are $(\langle k \rangle,\beta_1,\rho_{\texttt{ARV}})=(3.0,0.10,0)$.}
 \end{center}
\end{figure}

\begin{figure}
  \begin{center}
       \center \includegraphics[width=13cm]{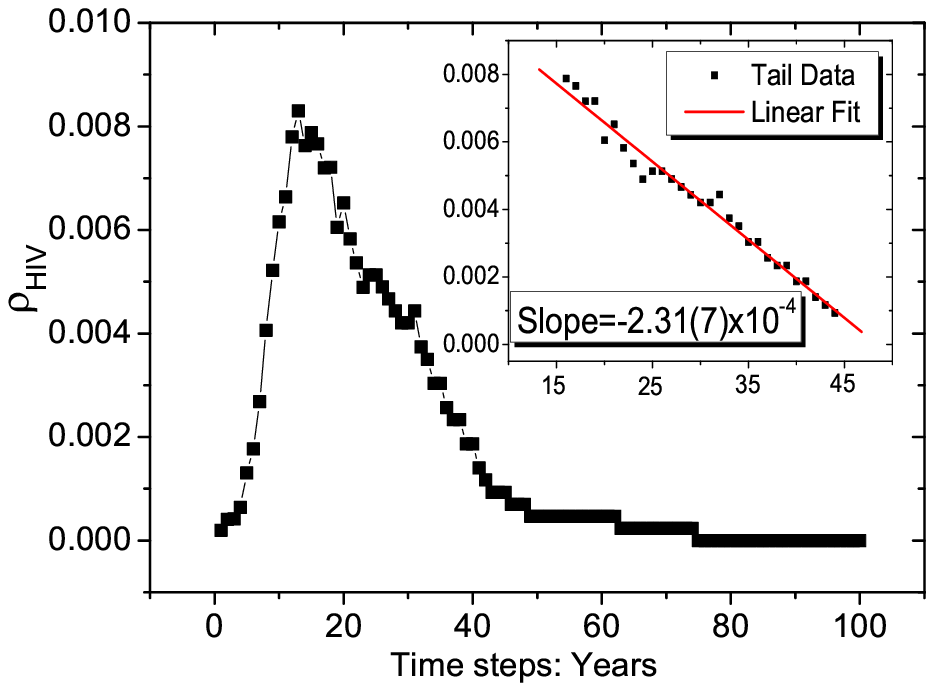}
       \caption{(Color Online) The prevalence of HIV verse time in a homogeneous network. The inset shows the tail of $\rho_{\texttt{ARV}}$ as a function of time step (from the peak point to nearly zero). The corresponding
       parameters are $(\langle k \rangle,\beta_1,\rho_{\texttt{ARV}})=(3.0,0.10,0)$.}
 \end{center}
\end{figure}

\subsection{Effect of the Heterogeneous Degree Distribution}
Many infections including HIV/AIDS can persistently exist in
population despite of a very low prevalence. This epidemiological
phenomenon can not be illuminated by previous models with
homogeneous mixing hypothesis \cite{Review3,Review4}. By using the
epidemic contact network with power-law degree distribution, the
present model can reproduce the above observed phenomenon, which is
in accordance with some previous theoretical studies about SIS/SIR
models on scale-free networks
\cite{Pastor-Satorras2001,May2001,Moreno2002,Gallos2003}. Note that,
since there are newly added susceptible individuals at each time
step, the dynamic behaviors of present model may be closer to SIS
model than SIR model. Figure 1 reports a typical simulation result
wherein the prevalence of HIV is only about $2\times10^{-3}$.
However, the infections can persistently exist for thousands years.
For comparison, we exhibit the situation under homogeneous mixing
hypothesis in figure 2, where the three parameters $(\langle k
\rangle,\beta_1,\rho_{\texttt{ARV}})$ are the same but all the nodes
have fixed degree 3. The prevalence increases in the early stage
since only few HIV-positive persons die, and then dies out obeying a
linear form.

In addition, this model displays oscillatory behaviors, which have
been observed in real world \cite{Cliff1984,Tohamsen1996} and
reproduced by some previous network epidemic models based on
small-world networks \cite{Kuperman2001,Xiong2004,Verdasca2005} or
scale-free networks \cite{Hayashi2004}.  One can see references
\cite{Watts1998,Watts1999} for the concept of small-world
networks. However, since the time from first report about HIV
cases to now is relative short compared with the oscillatory
period, we can not make sure if the real-life HIV epidemics
showing some kinds of oscillation.

\begin{figure}
  \begin{center}
       \center \includegraphics[width=13cm]{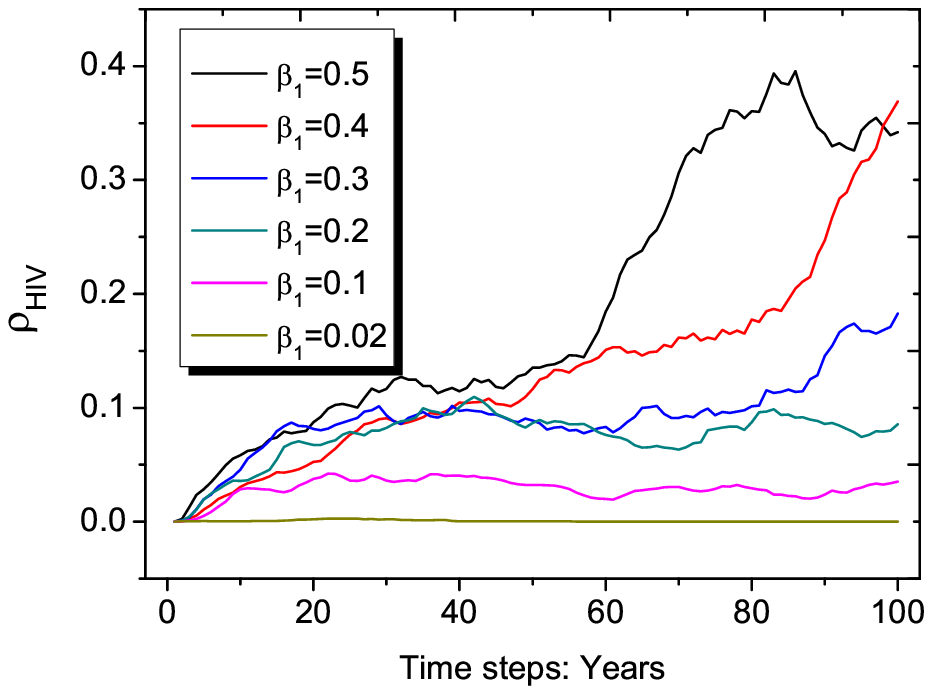}
       \caption{(Color Online) The prevalence of HIV verse time for different transmission probability. The corresponding parameters are  $(\langle k \rangle,\rho_{\texttt{ARV}})=(3.0,0)$.}
 \end{center}
\end{figure}

\begin{figure}
  \begin{center}
       \center \includegraphics[width=13cm]{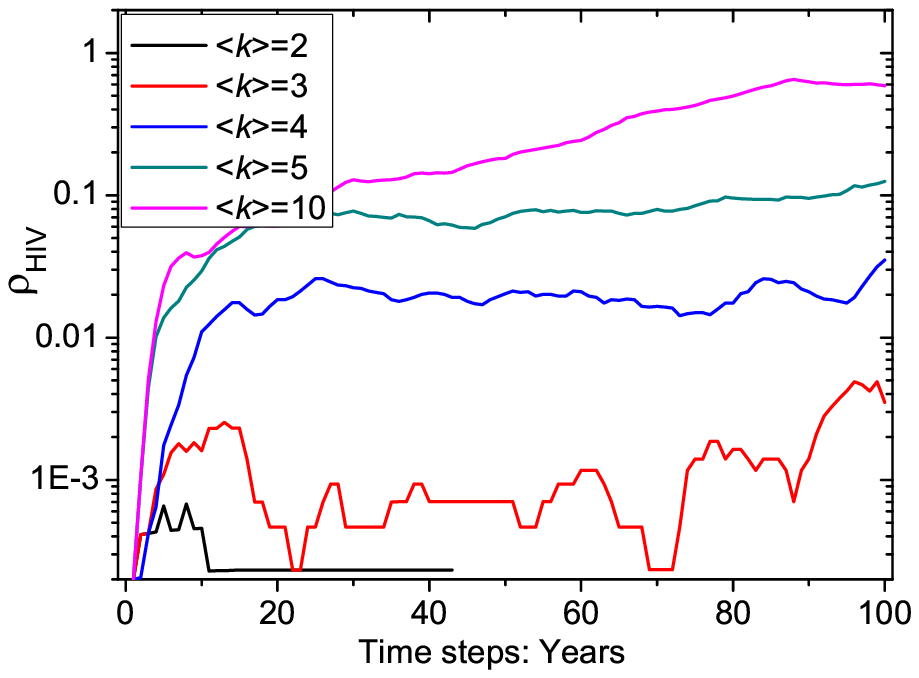}
       \caption{(Color Online) The prevalence of HIV verse time for different average degrees of epidemic contact networks. The corresponding parameters are  $(\beta_1,\rho_{\texttt{ARV}})=(0.10,0)$.}
 \end{center}
\end{figure}

\subsection{Effect of Transmission Probability}
The transmission probability $\beta_1$ not only depends on the
pathological characters of HIV, but only can be managed by
government and other organizations. For example, the popularization
of the usage of condoms will sharply reduce the transmission
probability per sexual partner/act as observed in Thailand and
Cambodia \cite{Cohen2003,Buckingham2004}. Figure 3 exhibits the
$\rho_{\texttt{ARV}}-t$ curves for different $\beta_1$: When
$\beta_1$ is large, the prevalence fleetly increase until
considerable ratio of whole population gets infected, while for
smaller $\beta_1$, the infection either persistently exists in a low
prevalence-level, or vanishes.

\subsection{Effect of Average Degree}
We have also investigated the effect of average degree $\langle k
\rangle$ on network epidemic behaviors. As shown in figure 4, the
behaviors of this model are very sensitive to the mean degree.
Clearly, larger mean degree will statistically enlarge the
probability of coming into contact with infected individuals, thus
leading to more serious situation. Combine this result and that of
\textbf{section 3.1}, one will find that the epidemic behaviors are
highly affected by the network topology.

\begin{figure}
  \begin{center}
       \center \includegraphics[width=13cm]{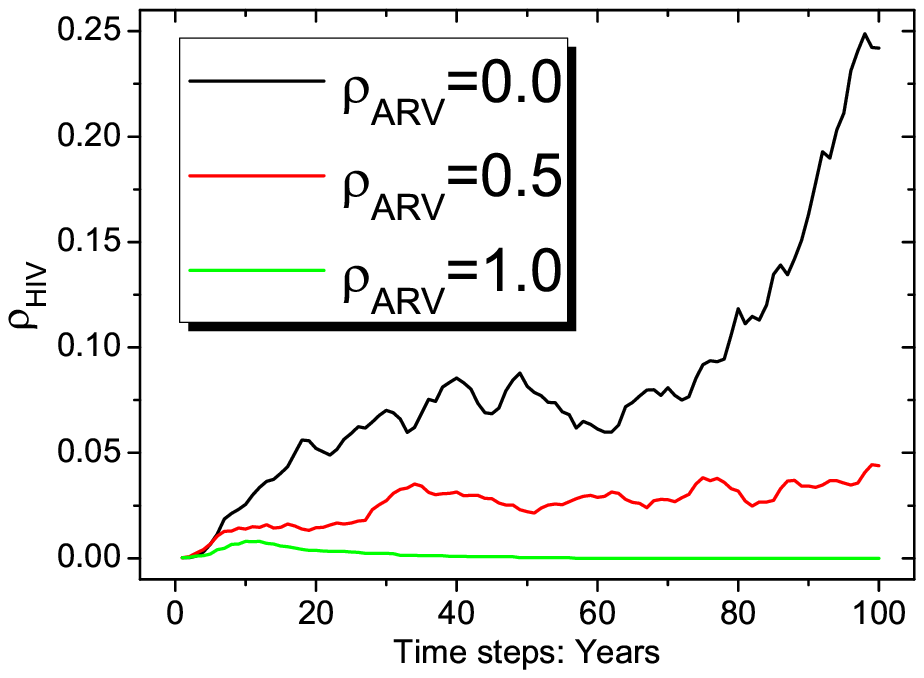}
       \caption{(Color Online) The prevalence of HIV verse time for different ARV treatment levels. The corresponding parameters are  $(\langle k \rangle, \beta_1)=(3,0.23)$. This value of $\beta_1$ is chosen according to the case of northern Thailand \cite{Nagachinta1997}.}
 \end{center}
\end{figure}

\begin{figure}
  \begin{center}
       \center \includegraphics[width=13cm]{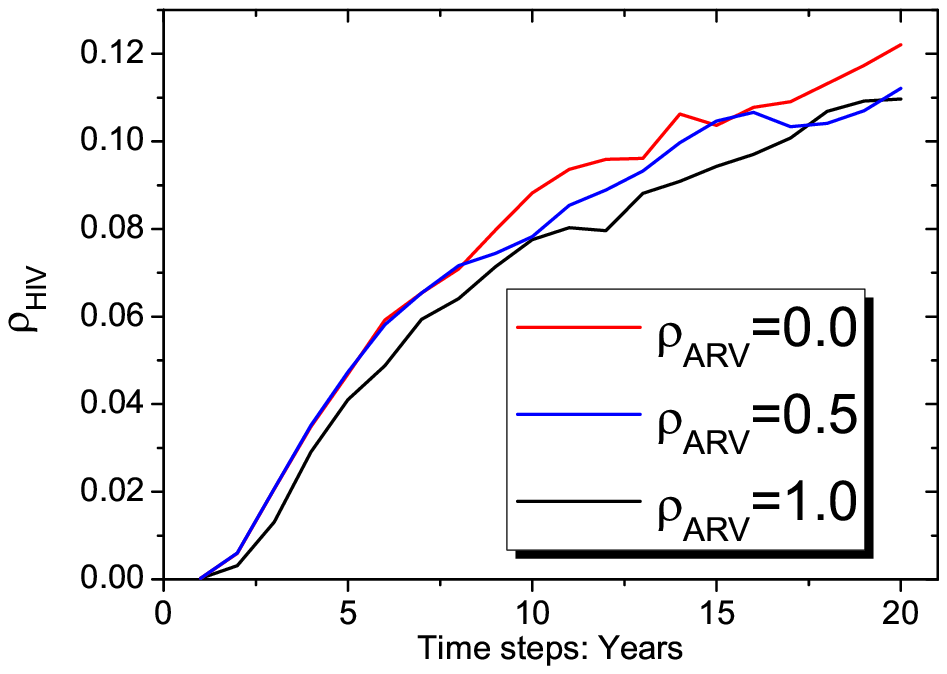}
       \caption{(Color Online) The prevalence of HIV verse time for different ARV treatment levels. The corresponding parameters are  $(\langle k \rangle, \beta_1)=(5,0.23)$.}
 \end{center}
\end{figure}

\subsection{Impact of Antiretroviral Drug Therapies}
The antiretroviral drug therapies have two opposite effects. On one
hand, it will reduce the probabilities of both sexual transmission
and perinatal transmission, thus ought to be very helpful in
controlling the epidemic spreads \cite{Porco2004,Sabin2005}. On the
other hand, this treatment will increase the life expectancy for
HIV-positive persons and these ARV users can infected more
individuals if they do not stop their risky behaviors, thus this
treatment may on the contrary increase the incidence of HIV/AIDS
\cite{Blower2000,Velasco-Hernandez2002}. Here, we assume the usage
of ARV will not change patients' behaviors, and in figure 5, one can
find that this treatment can substantially reduce HIV epidemics, and
even be possible to eradicate high prevalence HIV epidemics under
certain ideal conditions. It is worthwhile to emphasize that, the
simulation results in figure 5 strongly depend on the choices of
some dubious and imprecise parameters \cite{Porco2004,Sabin2005}
such as the ratios $\beta_2/\beta_1$, $\zeta_2/\zeta_1$ and
$\varepsilon_2/\varepsilon_1$. Therefore, the corresponding results
are not confessed. Maybe further empirical and experimental studies
about antiretroviral drug therapies may lead to more accurate
results. In addition, the behavior parameters $\langle k \rangle$
and $\beta_1$ are more important than $\rho_{\texttt{ARV}}$, and for
very large $(\langle k \rangle, \beta_1)$ case, the impact of
antiretroviral drug therapies is very weak as shown in figure 6.
Hence, to reduce the risky behaviors is much more effective in the
fight against HIV/AIDS rather than ARV treatment, especially for the
poor countries. The cases of Thailand \cite{Cohen2003} and Eastern
Zimbabwe are very good examples \cite{Gregson2006}.

\begin{figure}
  \begin{center}
       \center \includegraphics[width=13cm]{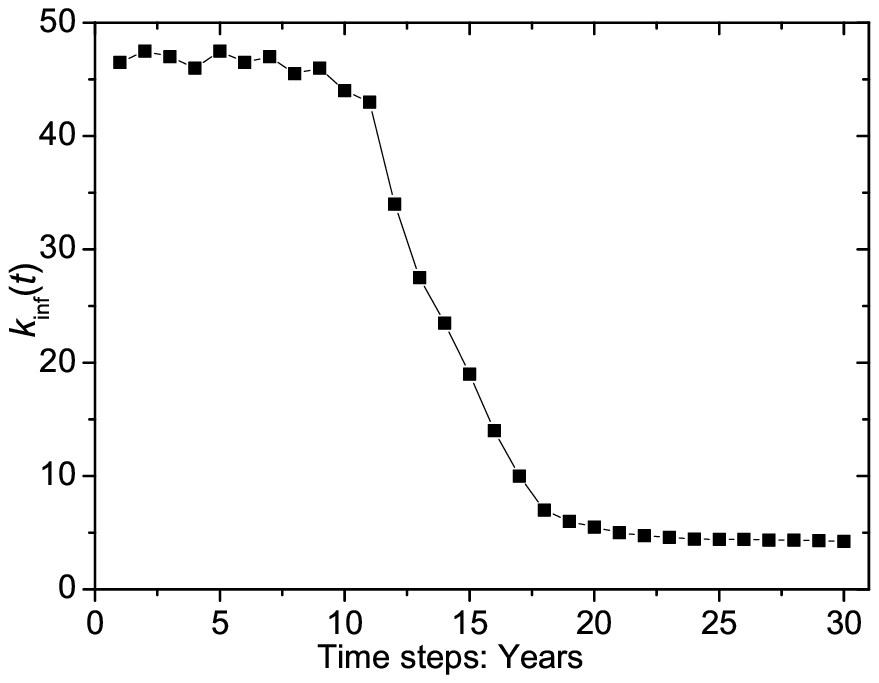}
       \caption{Time behavior of the average degree of the newly infected nodes. All the data are the average over ten realizations. The corresponding
       parameters are $(\langle k \rangle,\beta_1,\rho_{\texttt{ARV}})=(3.0,0.10,0)$.}
 \end{center}
\end{figure}

\subsection{Hierarchical Spread of HIV Epidemic Outbreaks}
In the epidemic contact networks, the degree can reflect the
susceptibility of individual to some extent, that is, the node with
higher degree is easier to be infected statistically
\cite{Christley2005}. Here, we investigate the behavior of the
average degree of the newly infected nodes in networks at time $t$,
denoted by $k_{inf}(t)$. We use the average of 10 realizations to
reduce the fluctuations. As shown in figure 7, the dynamical
spreading process is therefore clear: After the high-risk population
are infected within a short time, the spread is going towards
generic population (low-risk population). This hierarchical spread,
has been reported in some previous pure theoretical studies on SI
model \cite{SI1,SI2,SI3,SI4}, but not been emphasized in previous
HIV/AIDS epidemic models. However, this phenomenon has been observed
in real-life HIV epidemics: In the early stage the infection is
adhered to these high-risk persons, such as sex workers, injection
drug users, men who have sex with men, and so on. And then, it
diffuses to generic population. As an typical example, one can see
the situation in China \cite{CHINAIDS}.

\begin{figure}
  \begin{center}
       \center \includegraphics[width=14.5cm]{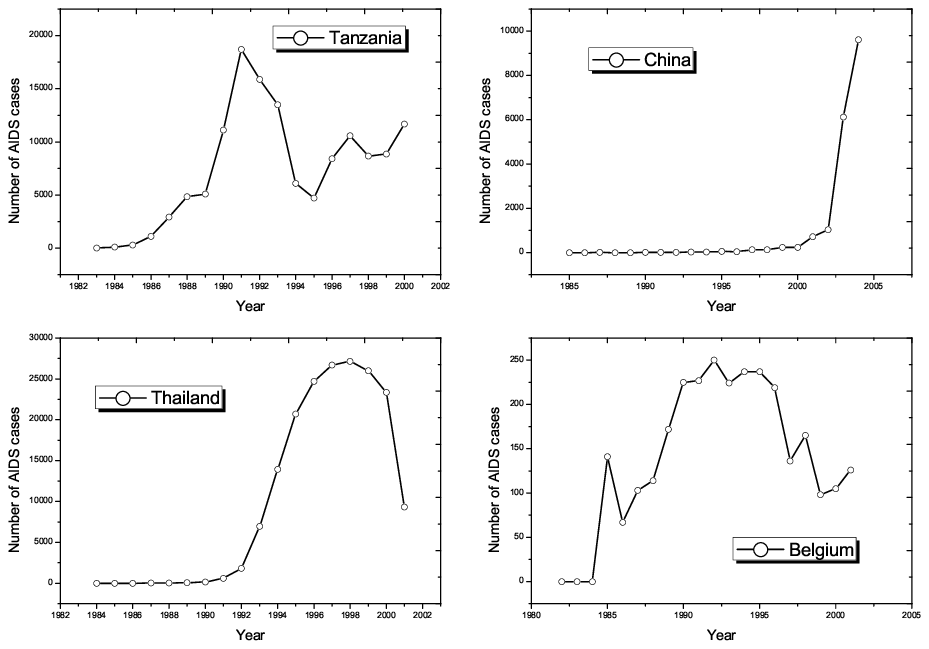}
       \caption{The time series of AIDS-patients number from national sentinel surveillances in Tanzania, China, Thailand and Belgium. These data are obtained form the web site of UNAIDS.}
 \end{center}
\end{figure}

\begin{figure}
  \begin{center}
       \center \includegraphics[width=6.5cm]{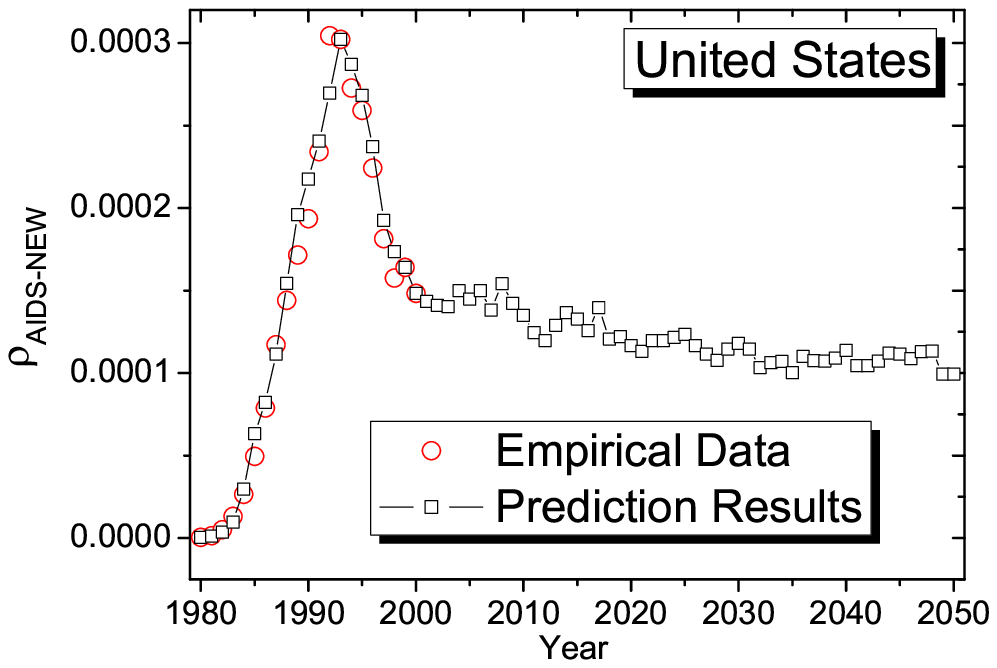} \includegraphics[width=6.5cm]{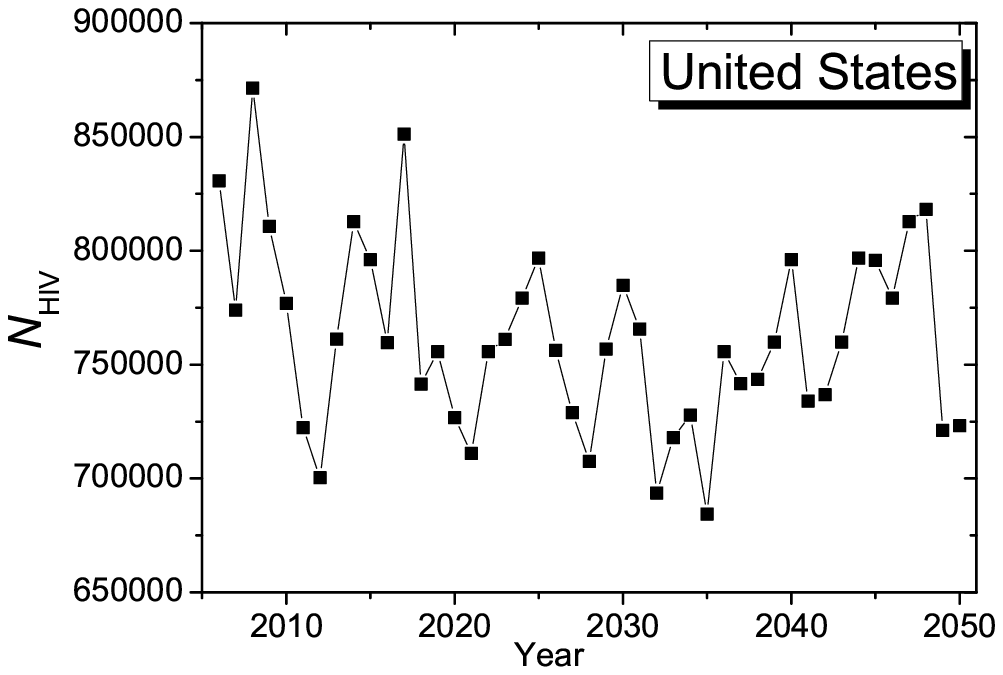}
       \caption{(Color Online) The prediction results for US. The left plot shows the empirical data and the projections about AID cases reported annually, and the right plot is the corresponding HIV-positive numbers from year 2006 to 2050. In this case, $\rho_{\texttt{ARV}}=0.65$, and the optimal parameters in this case are $\beta_1=0.11$ and $\langle k \rangle=4.3$.}
 \end{center}
\end{figure}

\section{Relevance to the real HIV/AIDS Epidemics}
Previous studies on the prediction of the HIV/AIDS epidemics
mainly concentrate on the data of the number of reported
HIV-positive cases. These methods, like empirical Bayesian
back-calculation method \cite{Downs1997}, can give a relatively
accurate prediction in short-term. However, it can not provide
useful information about the underlying dynamic mechanism.
Therefore, In this section, we will try to predict HIV/AIDS
epidemics by using the present model.

The lack of comprehensive and authentic data is one of the most
serious problems in evaluating and predicting HIV/AIDS epidemics.
For example, in the year 2004, the Chinese Minister of Health
reported that the number of living HIV-positive persons is about
$8.6\times 10^5$, but in the year 2006, it says that this number is
completely incorrect due to the greatly overvaluing. Actually, the
veracity of the reported HIV-positive numbers is dubious. From the
web site of UNAIDS \cite{UNAIDS}, except the data of HIV-positive
numbers, one can also obtain the data about the number of
AIDS-patients from national sentinel surveillances. These data are
also dubious since the monitor policies are not professional
especially in developing countries and some AIDS-patients do not
want to report to the sentinel surveillances. However, the data from
national sentinel surveillances do not involve external estimating
algorithm, thus we believe they are at least more faithworthy than
the HIV-positive numbers.

In figure 8, we show four typical forms of the time series of the
number of AIDS cases. Although there may be some other forms, the
present fours are representative. The most serious country is
Tanzania, wherein a considerable ratio of whole population is
infected. Without impelling control policies, Tanzania will be
completely destroyed before long. In China, the proportion of AIDS
cases seems very small but the amount of AIDS is quite large as a
result of the striking huge ensemble, and its quick and monotone
increasing trend brings us heavy misgivings. Thailand is a
successful example of external control. Once, Thailand, especially
the northern Thailand, is the most serious country in Asia due to
its thriving and prosperous pornographic business. Delightfully, the
government is cognizant of this problem and forces all the sex
workers using condoms. This policy leads to a sharply decreasing of
HIV-positive and AIDS-patient numbers. Some other countries, like
Brazil, have also achieved successful policies in controlling
HIV/AIDS epidemics. However, these emergent external policies bring
great challenges in predicting. The most optimistic situation is
that of Belgium. The HIV/AIDS persists in a very low prevalence
level and no increasing trend is observed.

In our model, according the assumption in \textbf{section 2.2}, we
consider the mortality at time $t$ as the number of newly monitored
AIDS-patients. This quantity, denoted by $N_{\texttt{AIDS-NEW}}$ ,
can be obtained from the model by combining the death rolls of
children, adults and old persons. Because of the computational
limit, we can at most handle the epidemic contact network with size
$N \sim 10^7$. However, the number of people aged from 15 to 49 in
some countries is much larger than $10^7$. In order to compare the
time series generated by our model and those of real country, all
the data are normalized by the population size aged from 15 to 49.
The normalized number of AIDS cases is denoted by
$\rho_{\texttt{AIDS-NEW}}$. In addition, we assume the number of
AIDS cases $N_{\texttt{AIDS-NEW}}$ is proportional to the
HIV-positive number $N_{\texttt{HIV}}$ at a given time $t$. Denote
the normalized data from sentinel surveillances $x(0),x(1),\cdots,
x(T)$, and the data generated from our model $y(0),y(1),\cdots,
y(T)$, the departure is defined as
\begin{equation}
e=\sum^T_{t=0}[x(t)-y(t)]^2.
\end{equation}
Since the parameter $\rho_{\texttt{ARV}}$ is known after the country
is selected (these data can also be obtained from UNAIDS), there are
only two tunable parameter $\beta_1$ and $\langle k \rangle$. Hence
this task degenerates to an optimal problem: Determine the proper
value of $\beta_1$ and $\langle k \rangle$ to minimize the departure
$e$. The optimal problem is carried out by searching all the values
of $(\beta_1,\langle k \rangle)$ in the Cartesian product of sets
$\{0.01,0.02,\cdots,0.50\}$ and $\{1.0,1.1,\cdots,10.0\}$, and
choosing the one corresponding to minimal $e$. The parameters will
not change with time, that is to say, the present prediction is
valid only for the cases with no additional interventions.

\begin{figure}
  \begin{center}
       \center \includegraphics[width=13cm]{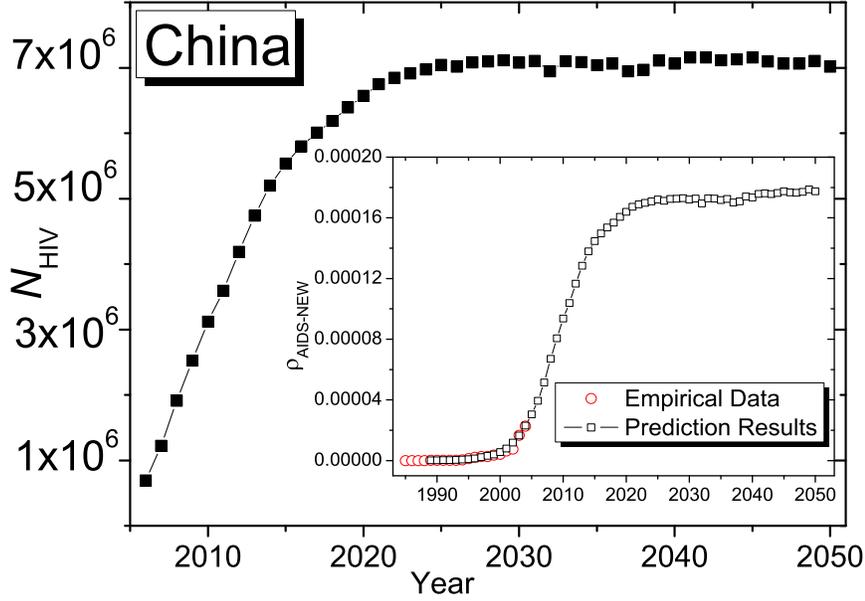}
       \caption{(Color Online) The prediction results for China. The main plot shows the empirical data and the projections about AID cases reported annually, and the inset is the corresponding HIV-positive numbers from year 2006 to 2050. In this case, $\rho_{\texttt{ARV}}=0.05$, and the optimal parameters in this case are $\beta_1=0.28$ and $\langle k \rangle=2.1$.}
 \end{center}
\end{figure}

\begin{figure}
  \begin{center}
       \center \includegraphics[width=13cm]{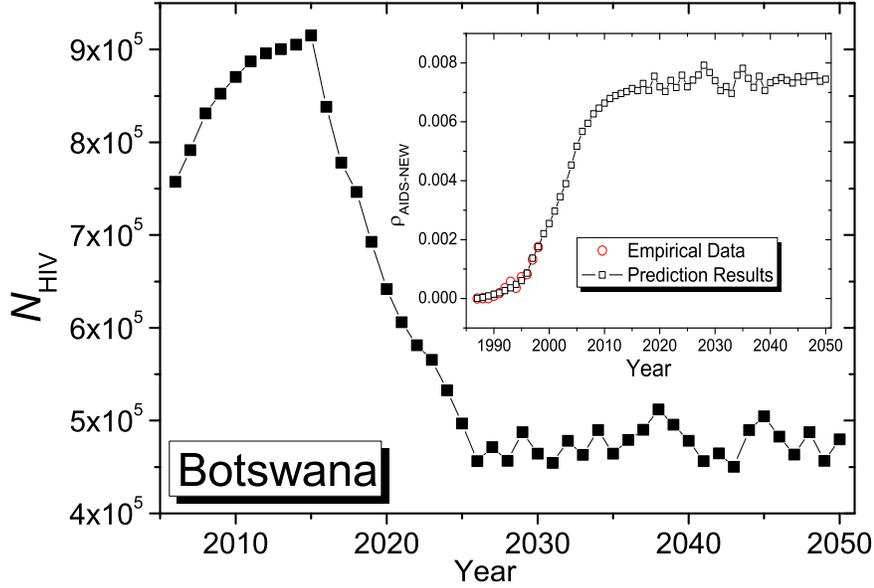}
       \caption{(Color Online) The prediction results for Botswana. The main plot shows the empirical data and the projections about AID cases reported annually, and the inset is the corresponding HIV-positive numbers from year 2006 to 2050. In this case, $\rho_{\texttt{ARV}}=0.079$, and the optimal parameters in this case are $\beta_1=0.31$ and $\langle k \rangle=7.2$.}
 \end{center}
\end{figure}

We have tried this prediction method for many representative
countries, and found the cases can be roughly divided into three
patterns. The typical example for the first pattern is the United
States (US), wherein the curve of AIDS-patient number has an obvious
peak before the year 2000, and then decreases to a relative low and
stable level. The similar behaviors have also been found for many
other countries, such as Mexico, Spain, Australia, Belgium,
Thailand, and so on. The common feature of these countries is that
their transmission probabilities are all small. This may be because
of the high popularization rates of condoms and disposable
injectors.

China is a particular example, although the prevalence of AIDS cases
are very low, it increases exponentially fast in the early stage
with exponent $\approx 0.15\pm0.01$, that is,
$\rho_{\texttt{AIDS-NEW}}\sim \texttt{e}^{0.15t}$ in the early
stage. However, this velocity will be slowed down and the number of
AIDS cases will get steady after the year 2025. This interesting
behavior may due to the particular values of $\beta_1$ and $\langle
k \rangle$ in China. Traditionally, Chinese women are not supposed
to have sex with a man other than their future spouse, thus the mean
degree is very small compared to 'Western-style' society. However,
since the popularization rate of the usage of condoms in China is
very low $(<20\%)$, the transmission probability in China is much
higher than these developed countries. In a word, although no
efficient policies in controlling HIV epidemics have been
implemented in Chinese government or other organizations, the
traditional moral sense may protect China from suffering AIDS.

Note that, the shapes of $N_{\texttt{HIV}}(t)$ and
$\rho_{\texttt{AIDS-NEW}}(t)$ are slightly different, which
attributes to the fluctuation of population size. Although the
numbers of HIV-positive dwellers seems high in some countries, such
as US and China, their direct and indirect effects on the
demographic structure of the whole population are very weak. If we
fixed $\beta=0$ for US and China from 2006 to 2050, then the
population size (aged from 15 to 49) without HIV/AIDS will be almost
the same as the original prediction results. Since the departures
can not be observed in plots, we have not shown here.

The most serious regions suffering AIDS are Africa (especially
Sub-Saharan Africa) \cite{Sinka2003,Lau2004,Hanson2005} as well as
Latin America and Caribbean region \cite{Calleja2002}. An typical
example is Botswana, which is a relative rich country in Africa but
has the highest HIV prevalence all over the world. If no additional
interventions, the infection will kill more than 50\% persons in
their prime of life. The ostensible stable behavior after 2020 is on
account of the existence of some no-risk people: They adhere to
monogamy and do not hit the pipe, thus will not be infected. In
network language, these individuals belong to some isolated
clusters. In network SIR model, these isolated clusters usually come
into being as a result of the removal of some individuals
\cite{Sander2002}. Other than US and China, demographic impact of
the HIV Epidemic in Botswana is striking. In figure 12, we compare
the predicted population size (aged from 15 to 49) with the no-AIDS
case where we set $\beta_1=0$ from 2006 to 2050. After year 2015,
the population size sharply declines, which is the very reason of
the decline of $N_{\texttt{HIV}}$ in figure 11. Only by ten years,
the population holds down to a half level. The demographic gets
stable in this new level since there are a number of no-risk people.
Finally, the age distribution in Botswana becomes sandglass-like
since many people in their prime of life (aged 15 to 49) will be
killed by AIDS. We are afraid that some other countries in Africa,
such as Malawi, Tanzania and Zambia, may face the same danger
\cite{Fylkesnes1998,Glynn2001}.

\begin{figure}
  \begin{center}
       \center \includegraphics[width=13cm]{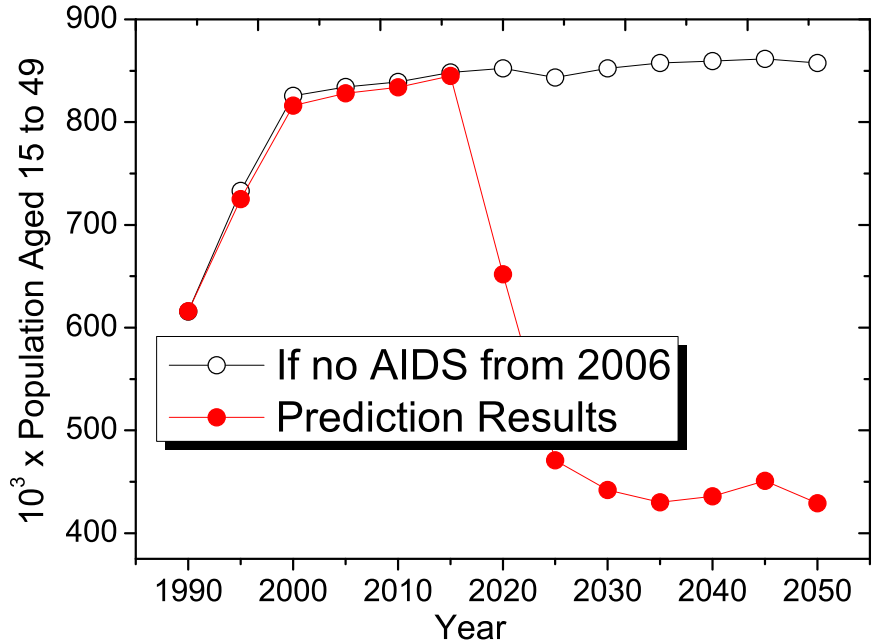}
       \caption{(Color Online) Predicted population size aged 15 to 49 comparing with the situation without AIDS.}
 \end{center}
\end{figure}

\section{Conclusion and Discussion}
In this article, we propose a network epidemic model for HIV
epidemics, wherein each individual is repreasented by a node of the
transmission network and the edges are the connections between
individuals along which the infection may spread. Motivated by some
previous empirical studies on the pattern of sexual contact, we set
the sexual activity of each individual, measured by its degree, is
not homogeneous but obeys a power-law distribution.

Many infections concluding HIV/AIDS can persist in population at a
very low prevalence. This epidemiological phenomenon can not be
illuminated by previous models with homogeneous mixing hypothesis,
while our model can reproduce this feature due to the
heterogeneity of activity. In addition, the model displays a clear
picture of hierarchical spread: In the early stage the infection
is adhered to these high-risk persons, and then, diffuses toward
low-risk population.

There are two main ingredients baffling the prediction obtained by
dynamical model: The first is the lack of comprehensive and
authentic data, and the second is the existence of unexpected
interventions like the governmental action against HIV/AIDS
epidemics. Right or wrong, we try to predict HIV epidemics by using
the present model, and hope it will at least capture some
qualitative features. The prediction results show that the
development of epidemics can be roughly categorized into three
patterns for different countries: persist in a stable and low level
after a peak in the early stage (US), monotonously grow and then
persist in a stable and low level (China), infect considerable ratio
of population. Which class the HIV epidemic of a given country
finally belongs to is mainly determined by the corresponding
behavior parameters $(\langle k \rangle,\beta_1)$. The interplay of
demographic structure and HIV epidemics is also taken into account.
In most cases, the effect of HIV epidemics on demographic structure
is very weak, while for some extremely countries, like Botswana, the
population size can be decpressed to a half, and the age structure
will become sandglass-like since many people in their prime of life
(from aged 15 to 49) will be killed by AIDS.

We believe this work may have some contribution in understanding the
underlying mechanism of HIV epidemic dynamics, since it can
naturally reproduce some important observed characters in HIV spread
that has not been emphasized in the previous models. However, is has
many shortages which should be adverted to and may be considered in
the future works. The first is the memory-limitation and
time-complexity in simulation block the directly studies on very
large systems. Therefore, we have to use the normalization method to
mimic the real countries with huge population. This size effect may
bring additional error in prediction. A recently proposed fast
algorithm \cite{Moreno2003,Moreno2004} may improve the situation if
we have successfully modified some dynamical rules and translated
this model into an equal rate-equation form. Secondly, this model
consider only the heterosexual contacts and perinatal transmission,
however, other transmission routes, especially the homosexual
contacts \cite{Amirkhanian2001,Wade2005} and sharing injectors among
drug users \cite{Chu2005,Mocoy2004}, are also significant in HIV
epidemics. Finally, some details are ignored. For example, a recent
study \cite{Lewis2004} indicates the existence of large fertility
differentials between HIV-infected and uninfected women, and some
empirical studies show that the social networks have community
structure \cite{Girvan2002,Palla2005}, which will affect the
epidemic dynamics \cite{Liu2005,Yan2006}.

\section{Acknowledgments}
This work has been partially supported by the National Natural
Science Foundation of China under Grant Nos. 70471033, 10472116,
10532060, 70571074 and 10547004, the Specialized Research Fund for
the Doctoral Program of Higher Education (SRFDP No.20020358009),
the Special Research Founds for Theoretical Physics Frontier
Problems under Grant No. A0524701, and Specialized Program under
President Funding of Chinese Academy of Science.

\section{Appendix A: The Population and Demographic Data}
All the population and demographic data come from the United Nations
and can be obtained by decompressing the .zip file from
Http://www.comap.com/
undergraduate/contests/mcm/2006/problems/2006\%20ICM.zip. The
explanation about some used data in this article are as following.

\textbf{age\_data.xls}: These data give population (for both
sexes, and by gender) by five-year groups, major area, region, and
country, 1950-2050 (a. Estimates for 1950-2005; b. Projections for
2010-2050).

\textbf{fertility\_data.xls}: These data give age-specific
fertility rates by major area, region, and country, 1995-2050 (a.
Estimates for 1995-2005; b. Projections for 2010-2050).

\textbf{population\_data.xls}: These data give total population
(both sexes combined) by major area, region, and country, annually
for 1995-2050 (a. Estimates for 1950-2005; b. Projections for
2006-2050).

\end{document}